\documentclass[a4paper]{revtex4}
\usepackage{graphicx}
\usepackage{fancyhdr}
\usepackage{amsmath,amssymb,color}
\newcommand{\red}[1]{{\color[rgb]{1,0,0} #1}}
\newcommand{\U}{{\text{U}}}
\newcommand{\SU}{{\text{SU}}}
\newcommand{\eV}{{\text{eV}}}
\newcommand{\GeV}{{\text{GeV}}}

\newcommand{\BR}{{\text{BR}}}

\usepackage{color}

\begin{document}

\title{Radiative Neutrino Mass Models}

%

\author{Hiroaki Sugiyama}
\affiliation{
Maskawa Institute for Science and Culture,
Kyoto Sangyo University, Kyoto 603-8555, Japan
\footnote{
 Current affiliation is
Department of Physics, University of Toyama, Toyama 930-8555, Japan.
}
}

\begin{abstract}
 In this short review,
we see some typical models
in which light neutrino masses
are generated at the loop level.
 These models involve new Higgs bosons
whose Yukawa interactions with leptons
are constrained by the neutrino oscillation data.
 Predictions about flavor structures
of $\ell \to \overline{\ell}_1\ell_2\ell_3$
and leptonic decays of new Higgs bosons
via the constrained Yukawa interactions
are briefly summarized
in order to utilize such Higgs as
a probe of $\nu$ physics.

\end{abstract}

\maketitle

\thispagestyle{fancy}


\section{Introduction}

 Properties of the discovered Higgs boson~\cite{LHC-Higgs}
tells us about the mechanism
to generate particle masses~\cite{coupling}.
 Since neutrino masses
are extremely smaller than
masses of the other fermions in the standard model,
there would be
a new mechanism specific for generating neutrino masses.
 In such a new mechanism for neutrino masses,
it seems natural to expect that
there are some new Higgs bosons
relevant to the mechanism.
 Then,
we can utilize such new Higgs bosons
as a probe of $\nu$ physics.

 If we restrict ourselves to use the fields
which exist in the standard model,
the light neutrino mass $m_\nu^{}$ comes from
a dimension-5 operator
$(\overline{L^c} \epsilon \Phi) (\Phi^T \epsilon L)/\Lambda$,
where $\epsilon$ is the $2\times2$ completely antisymmetric tensor,
$L$ is an $\SU(2)_L$-doublet of leptons,
and $\Phi$ denotes the $\SU(2)_L$-doublet scalar field
in the standard model,
and $\Lambda$ is the energy scale of the new physics%
~\footnote
{
 Other two operators
$(\overline{L^c} L) (\Phi^T \Phi)/\Lambda$
and $(\overline{L^c} \Phi) (\Phi^T L)/\Lambda$
are also allowed.
 Each of three operators can be rewritten as
a linear combination of the others
via the Fierz transformation.
}.
 The seesaw mechanism~\cite{Ref:Type-I} is
the most familiar one to generate the dimension-5 operator
at the tree level.
 However,
the mechanism does not seems testable
because the suppression of the neutrino masses
is achieved by introducing extremely heavy
right-handed Majorana neutrinos.
 For example,
$\Lambda \sim 10^{15}\,\GeV$
for the right-handed Majorana neutrino mass
gives $m_\nu^{} \sim 0.1\,\eV$.
 On the other hand,
such a dimension-5 operator
can be obtained at the $n$-loop level
with an extra suppression factor of $(1/16\pi^2)^n$.
 For example,
$\Lambda \sim 10^3\,\GeV$ for $n=5$
can give $m_\nu^{} \sim 0.1\,\eV$.
 Even for a smaller $n$,
the neutrino mass can be sufficiently suppressed
with $\Lambda \sim 10^3\,\GeV$
because $m_\nu^{}$ can be suppressed
also by a product of new coupling constants
(each of them would be much less than unity)
which appear in the loop diagram.
 Thus,
new particles in such models of the radiative neutrino mass
could be observed at collider experiments.

 In radiative neutrino mass models,
new scalar fields are always added to the standard model,
and matrices of their Yukawa coupling constants
determine the structure of the neutrino mass matrix.
 Inversely,
new Yukawa matrices can be constrained
by the structure of the neutrino mass matrix
which is determined by the neutrino oscillation data.
 The constrained Yukawa matrices
give predictions about flavor structures of
$\ell \to \overline{\ell}_1\ell_2\ell_3$
and decays of new scalar particles into charged leptons.
 In this review,
we summarize what kinds of new particles
are introduced in some typical models
of the radiative neutrino mass.
 Then,
we see predictions
about these processes
in order to utilize them
for the test of these models.

\section{Models}

\renewcommand{\arraystretch}{1.5}
\begin{table}[t]
\begin{center}
\begin{tabular}{|r|rc|c||ccc|cc|ccc|c|cc|p{5mm}|}
\hline
\multicolumn{4}{|r||}{}
 & \multicolumn{3}{|c|}{$\psi_R^0$}
 & \multicolumn{2}{|c|}{$s^0$}
 & \multicolumn{3}{|c|}{$s^+$}
 & $s^{++}$
 & \multicolumn{2}{|c|}{$\Phi_2$}
 & \multicolumn{1}{|c|}{$\Delta$}\\
\hline
\multicolumn{4}{|r||}{Spin \ }
 & \multicolumn{3}{|c|}{$1/2$}
 & \multicolumn{2}{|c|}{$0$}
 & \multicolumn{3}{|c|}{$0$}
 & $0$
 & \multicolumn{2}{|c|}{$0$}
 & \multicolumn{1}{|c|}{$0$}\\
\hline
\multicolumn{4}{|r||}{$\SU(2)_L^{}$ \ }
 & \multicolumn{3}{|c|}{$\underline{\bf 1}$}
 & \multicolumn{2}{|c|}{$\underline{\bf 1}$}
 & \multicolumn{3}{|c|}{$\underline{\bf 1}$}
 & $\underline{\bf 1}$
 & \multicolumn{2}{|c|}{$\underline{\bf 2}$}
 & \multicolumn{1}{|c|}{$\underline{\bf 3}$}\\
\hline
\multicolumn{4}{|r||}{$\U(1)_Y^{}$ \ }
 & \multicolumn{3}{|c|}{$0$}
 & \multicolumn{2}{|c|}{$0$}
 & \multicolumn{3}{|c|}{$1$}
 & $2$
 & \multicolumn{2}{|c|}{$1/2$}
 & \multicolumn{1}{|c|}{$1$}\\
\hline\hline
\ Majorana $\nu$ \
 & \ Zee Model & \cite{Zee:1980ai} \ & \ 1-loop \
 & {} & {} & {}
 & {} & {}
 & {} & $\checkmark$ & {}
 & {}
 & \ $\checkmark$ & {}
 & {}\\
\cline{2-16}
{}
 & \ Zee-Babu Model & \cite{ZB} \ & \ 2-loop \
 & {} & {} & {}
 & {} & {}
 & {} & $\checkmark$ & {}
 & $\checkmark$
 & {} & {}
 & {}\\
\cline{2-16}
{}
 & \ Ma Model & \cite{Ma:2006km} \ & \ 1-loop \
 & {} & \red{$\dagger$} & {}
 & {} & {}
 & {} & {} & {}
 & {}
 & {} & \red{$\dagger$}
 & {}\\
\cline{2-16}
{}
 & \ Krauss-Nasri-Trodden Model & \cite{Krauss:2002px} \ & \ 3-loop \
 & {} & \red{$\dagger$} & {}
 & {} & {}
 & {} & $\checkmark$ & \red{$\dagger$} \
 & {}
 & {} & {}
 & {}\\
\cline{2-16}
{}
 & \ Aoki-Kanemura-Seto Model & \cite{AKS} \ & \ 3-loop \
 & {} & \red{$\dagger$} & {}
 & {} & \red{$\dagger$} \
 & {} & {} & \red{$\dagger$} \
 & {}
 & \ $\checkmark$ & {}
 & {}\\
\cline{2-16}
{}
 & \ Gustafsson-No-Rivera Model & \cite{Gustafsson:2012vj} \ & \ 3-loop \
 & {} & {} & {}
 & {} & {}
 & {} & {} & \red{$\dagger$} \
 & $\checkmark$
 & {} & \red{$\dagger$} \
 & {}\\
\cline{2-16}
{}
 & \ Kanemura-Sugiyama Model & \cite{Kanemura:2012rj} \ & \ 1-loop \
 & {} & {} & {}
 & {} & {}
 & {} & $\checkmark$ & \red{$\dagger$} \
 & {}
 & {} & \red{$\dagger$} \
 & \multicolumn{1}{|c|}{$\checkmark$}\\
\hline\hline
\ Dirac $\nu$ \
 & \ Nasri-Moussa Model & \cite{Nasri:2001ax} \ & \ 1-loop \
 & \ $\checkmark$ & {} & {}
 & {} & {}
 & \ $\checkmark$ & $\checkmark$ & {}
 & {}
 & {} & {}
 & {}\\
\cline{2-16}
{}
 & \ Gu-Sarkar Model & \cite{Gu:2007ug} \ & \ 1-loop \
 & \ $\checkmark$ & \red{$\dagger$} & \red{$\dagger$} \
 & \ $\checkmark$ & \red{$\dagger$} \
 & {} & {} & {}
 & {}
 & {} & \red{$\dagger$} \
 & {}\\
\cline{2-16}
{}
 & \ Kanemura-Matsui-Sugiyama Model & \cite{Kanemura:2013qva} \ & \ 1-loop \
 & \ $\checkmark$ & {} & {}
 & \ $\checkmark$ & \red{$\dagger$} \
 & {} & {} & {}
 & {}
 & \ $\checkmark$ & \red{$\dagger$} \
 & {}\\
\hline
\end{tabular}
\caption{
 A list of new particles introduced
in typical models of the radiative neutrino mass.
Red daggers~(\red{$\dagger$}) indicate that
these particles have the odd parity
under an unbroken $Z_2$ symmetry
(or charged under an unbroken global $\U(1)$ symmetry).
}
\label{tab:models}
\end{center}
\end{table}
\renewcommand{\arraystretch}{1}

 Let us briefly see some typical models
of the radiative neutrino mass.
 Particles introduced in these models
are summarized in Table~\ref{tab:models}.
 A checkmark~($\checkmark$)
or a red dagger~(\red{$\dagger$}) in the table means
that the particle is introduced in the model.
 The red dagger also shows
that the particle is odd under an unbroken $Z_2$ symmetry
(or charged under an unbroken global $\U(1)$ symmetry).
 Right-handed fermions $\psi_R^0$
stand for singlet fields under the standard model gauge group.
 Flavor indices of fermions are ignored for simplicity.
 Scalar fields of the $\SU(2)_L$-singlet representation
with hypercharges $Y=0$, $1$, and $2$ are
indicated by $s^0$, $s^+$, and $s^{++}$, respectively.
 The $\Phi_i$~$(i = 2, 3, \cdots)$
are $\SU(2)_L$-doublet scalar fields with $Y=1/2$
in addition to $\Phi_1$ in the standard model.
 The $\Delta$ mean $\SU(2)_L$-triplet scalar fields with $Y=1$.

\begin{figure}[t]
 \includegraphics[scale=0.45]{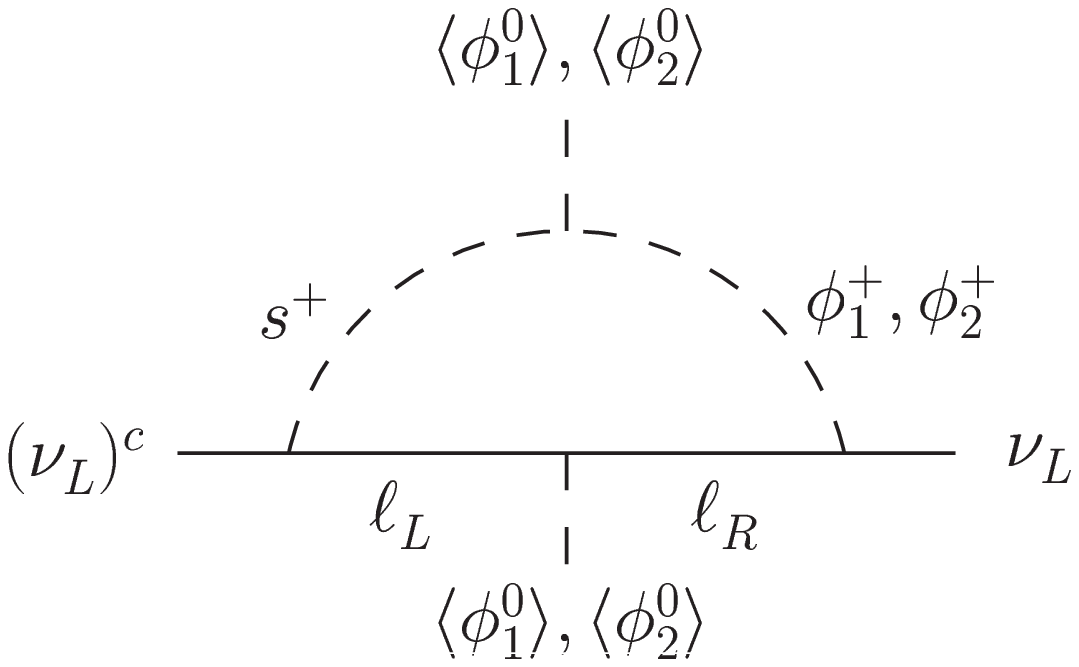}
 \includegraphics[scale=0.45]{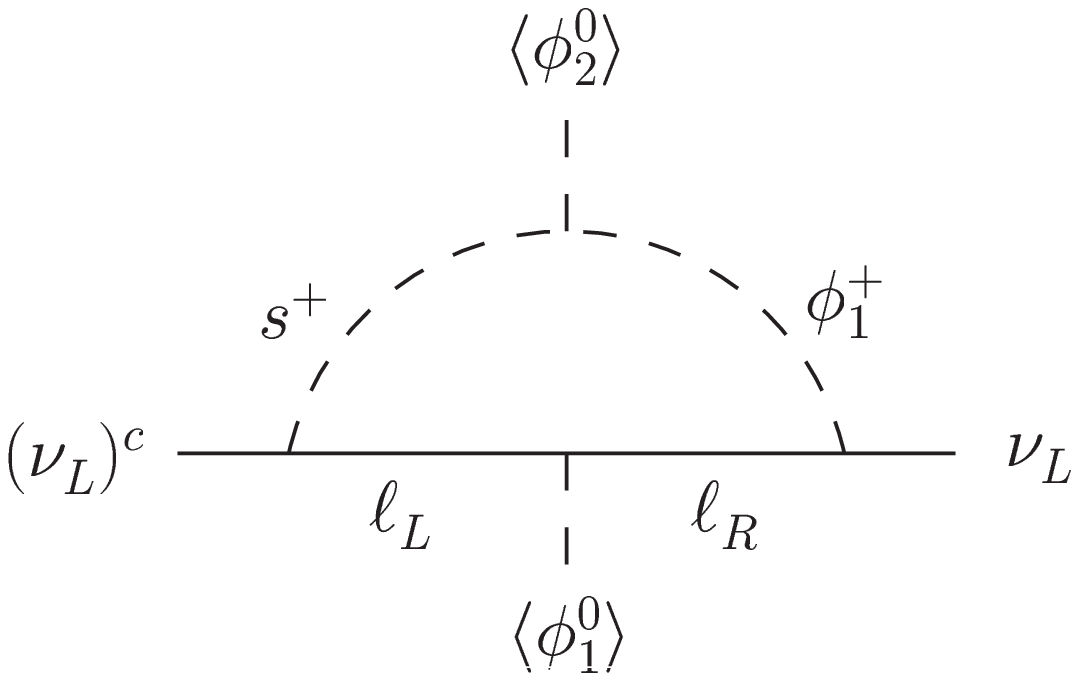}
 \vspace*{-2mm}
 \caption{
 The loop diagram for the Majorana neutrino mass in the Zee model~(left).
 The right diagram is a simplified one without FCNC\@.
 }
 \label{fig:Zee}
\end{figure}

 The {\bf Zee model}~\cite{Zee:1980ai}
is the first model
in which neutrino masses arise at the loop level.
 New particles introduced in the model are
$s^+$ and the second $\SU(2)_L$-doublet scalar field $\Phi_2$.
 Majorana neutrino masses are generated
by a sum of the 1-loop diagram in Fig.~\ref{fig:Zee}~(left)
and its transpose.
 The model can be simplified~\cite{Wolfenstein:1980sy}
such as each of fermions couples
only with one of two $\SU(2)_L$-doublet scalar fields%
~(Fig.~\ref{fig:Zee}~(right))
in order to forbid the flavor changing neutral current~(FCNC).
 Although
the simplified model~(the Zee-Wolfenstein model)
gives a predictive structure of the neutrino mass matrix,
the model was excluded~(See e.g., Ref.~\cite{He:2003ih})
by neutrino oscillation data.
 However, it should be noticed that
the original Zee model is still alive%
~(See e.g., Ref.~\cite{He:2011hs})
since we can accept the FCNC in the lepton sector.
 The structure of the neutrino mass matrix
is acceptable in the Zee model
even for a simplification with $m_e^{} = m_\mu^{} = 0$.

\begin{figure}[t]
\begin{minipage}{0.45\hsize}
\vspace*{4mm}
 \includegraphics[scale=0.45]{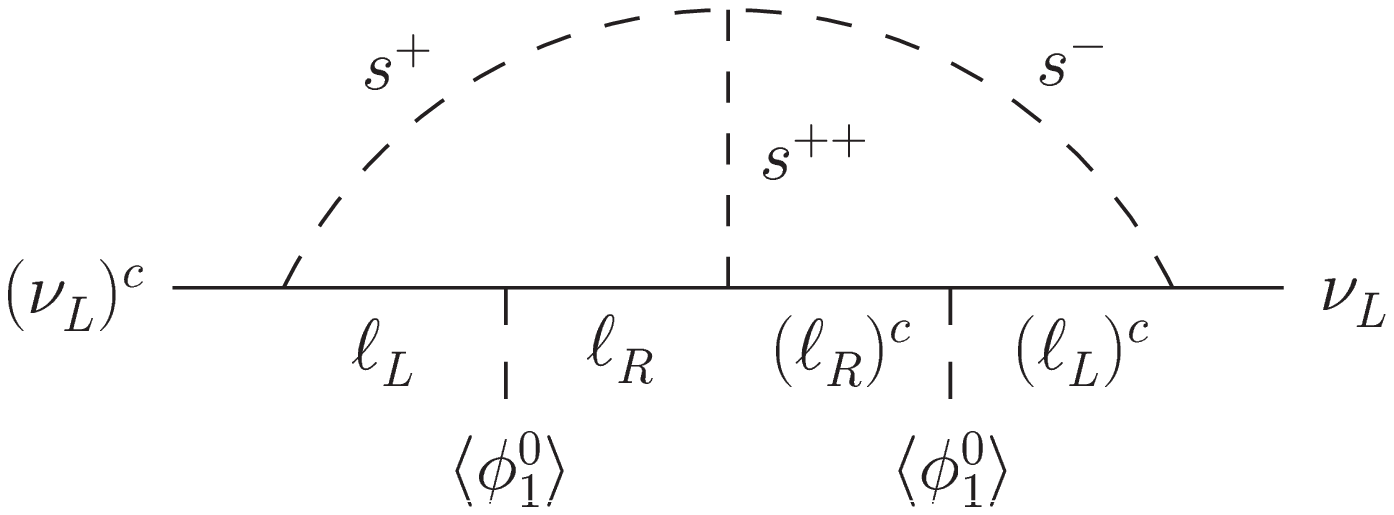}
 \vspace*{-2mm}
 \caption{
 The loop diagram for the Majorana neutrino mass in the Zee-Babu model.
 }
 \label{fig:ZB}
\end{minipage}
\qquad
\begin{minipage}{0.45\hsize}
 \includegraphics[scale=0.45]{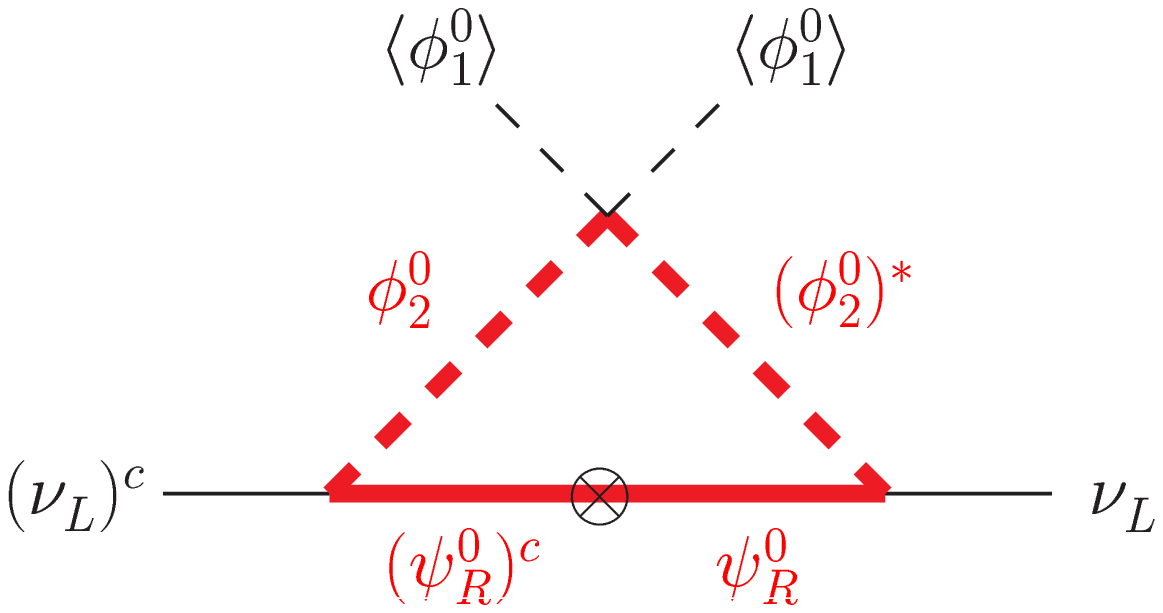}
 \vspace*{-2mm}
 \caption{
 The loop diagram for the Majorana neutrino mass in the Ma model.
 }
 \label{fig:Ma}
\end{minipage}
\end{figure}

 In the {\bf Zee-Babu model}~(ZB model)~\cite{ZB},
Majorana neutrino masses are generated
by the 2-loop diagram in Fig.~\ref{fig:ZB}
in which $s^+$ and $s^{++}$ are utilized.
 The scale of the neutrino mass $m_\nu^{}$
can be naively given by
$y_\text{new}^3\, \mu_3^{} m_\tau^2/(16 \pi^2 M)^2$,
where $y_\text{new}^{}$ denotes new Yukawa coupling constant
(a common value for $s^+$ and $s^{++}$ is assumed),
the coupling constant $\mu_3$~(its mass-dimension is 1)
is for the $s^+ s^+ s^{--}$ interaction,
and $M$ stands for the typical mass scale
of these new particles.
 Let's naively take the electroweak scale $O(100)\,\GeV$
for $\mu_3^{}$ and $M$,
which would enable us to
discover new scalar particles experimentally in the future.
 A naive expectation for the size of $y_\text{new}^{}$
would be the order of the Yukawa coupling constant
for the $\tau$ lepton, $O(10^{-2})$.
 Then,
the naive estimation in the ZB model gives
an appropriate neutrino mass scale $m_\nu^{} \sim 0.1\,\eV$.
 It is worth to mention that
the ZB model is viable for the model of the neutrino mass
even if $m_e$ is simply ignored.

 If a conserved charge is assigned only to new particles,
it is easy to construct a loop diagram
which involves only new particles in the loop.
 In addition,
the lightest one
among the charged particles becomes stable,
and the one can be a dark matter candidate
if it is electrically neutral.
 In the {\bf Ma model}~\cite{Ma:2006km}
as the simplest example,
$\psi_R^0$ and the second $\SU(2)_L$-doublet scalar field $\Phi_2$
are introduced such as
they have the odd parity~("charge" $-1$)
under an unbroken $Z_2$ symmetry
while the standard model particles have
the even parity~("charge" $1$).
 These $Z_2$-odd particles are utilized
in the 1-loop diagram~(Fig.~\ref{fig:Ma})
for Majorana neutrino masses.
 When $\psi_R^0$ or $\text{Re}(\phi_2^0)$ or $\text{Im}(\phi_2^0)$
is the lightest $Z_2$-odd particle,
the particle can be considered as a dark matter candidate.



\begin{figure}[t]
 \includegraphics[scale=0.5]{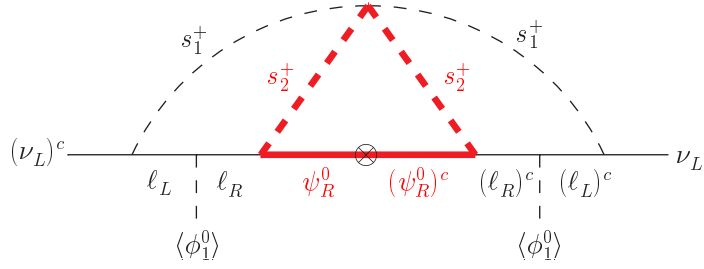}
 \vspace*{-2mm}
 \caption{
 The loop diagram for the Majorana neutrino mass
 in the Krauss-Nasri-Trodden model.
 }
 \label{fig:KNT}
\end{figure}

 The first model of the radiatively generated neutrino mass
with the dark matter candidate in the loop
is the {\bf Krauss-Nasri-Trodden model}~(KNT model)~\cite{Krauss:2002px}.
 Majorana neutrino masses come from
the 3-loop diagram~(Fig.~\ref{fig:KNT}).
 Fermions $\psi_R^0$ and an $\SU(2)_L$-singlet scalar field $s_2^+$
are introduced as $Z_2$-odd particles
while the other singlet scalar field $s_1^+$
and the standard model particles are $Z_2$-even ones.
 The $\psi_R^0$ is the dark matter candidate
if it is the lightest $Z_2$-odd particle.
 Similarly to the ZB model,
$m_e^{}$ can be ignored in the loop diagram.

\begin{figure}[t]
\begin{minipage}{0.45\hsize}
 \vspace*{5mm}
 \includegraphics[scale=0.45]{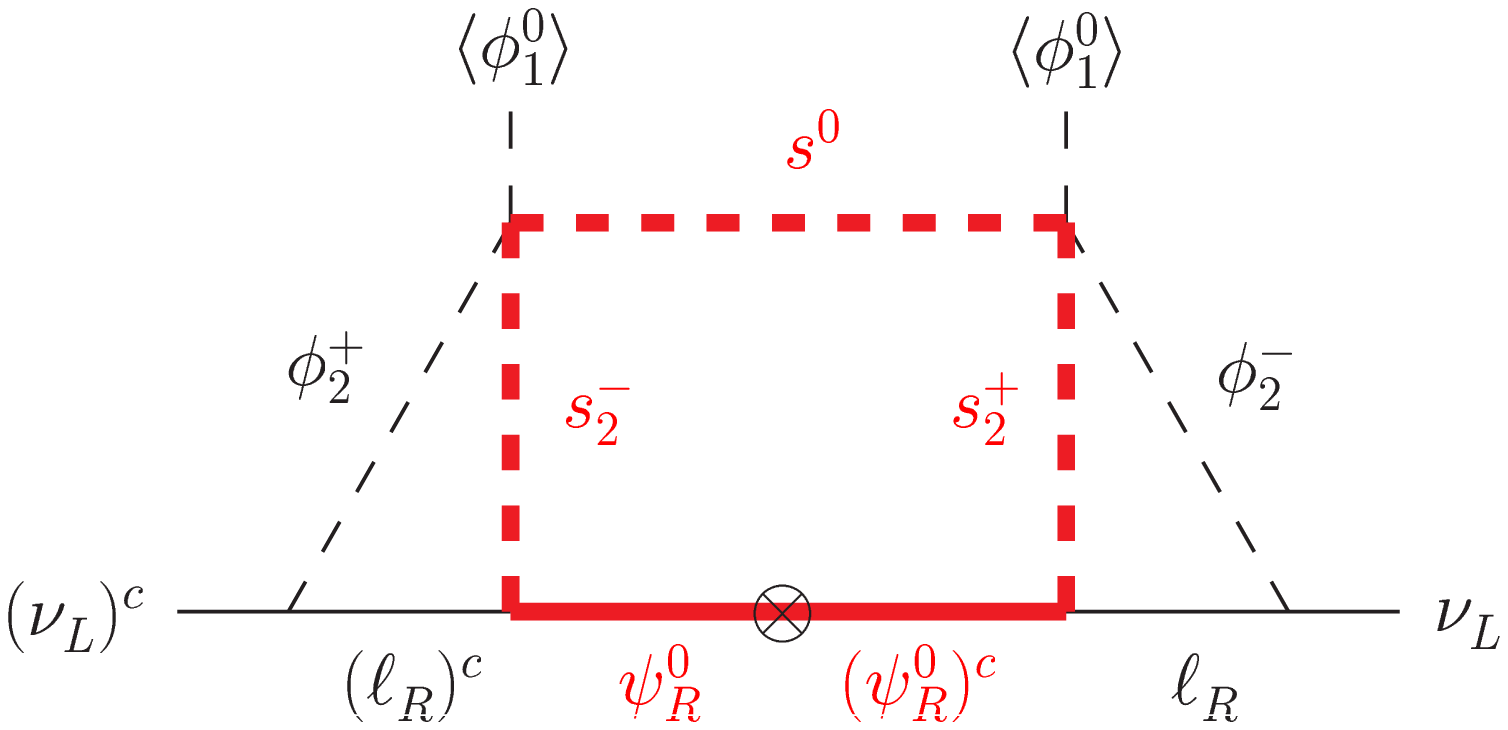}
 \vspace*{3mm}
 \caption{
 The loop diagram for the Majorana neutrino mass
 in the Aoki-Kanemura-Seto model.
 }
 \label{fig:AKS}
\end{minipage}
\qquad
\begin{minipage}{0.45\hsize}
 \includegraphics[scale=0.45]{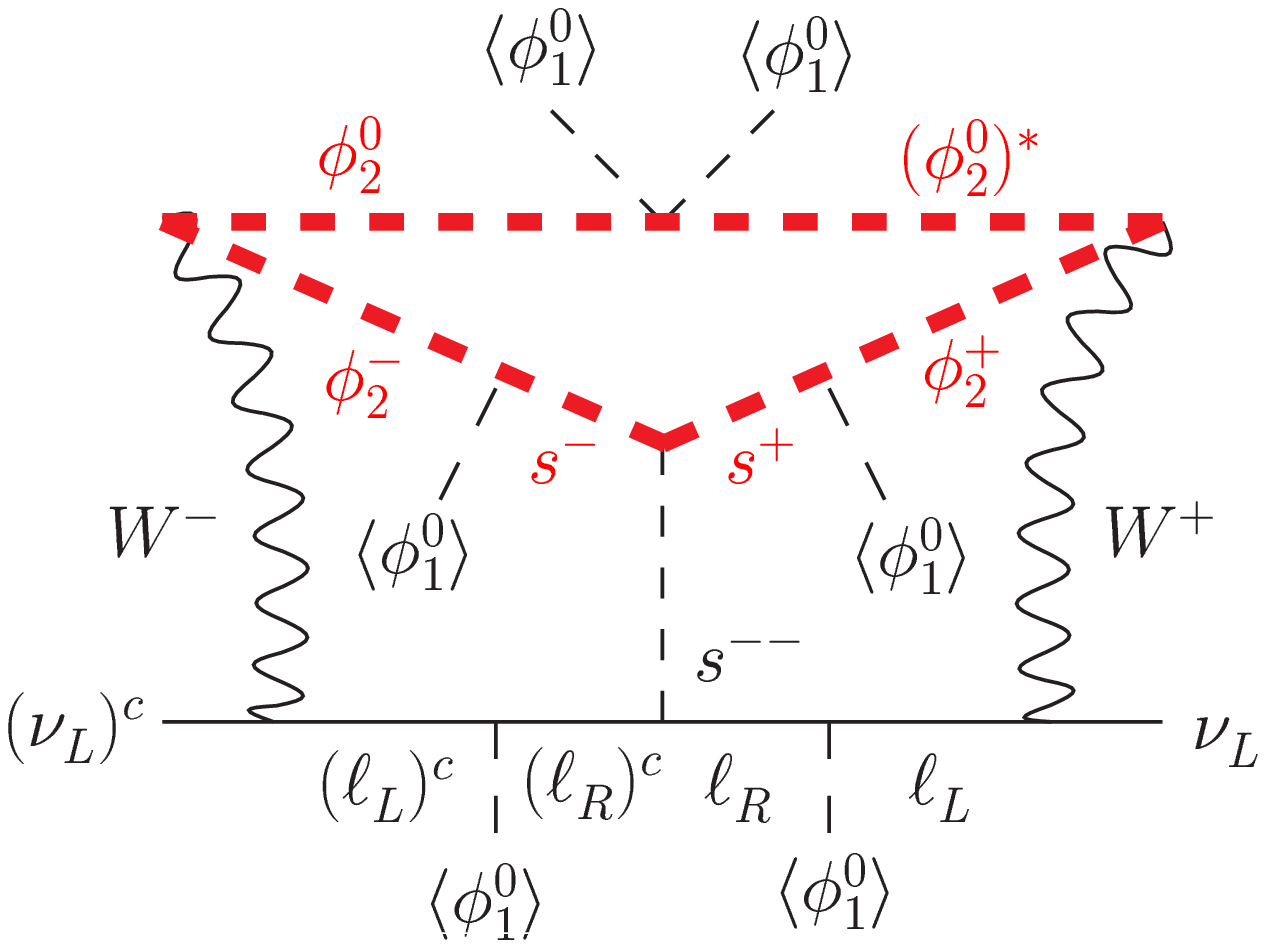}
 \vspace*{-2mm}
 \caption{
 The loop diagram for the Majorana neutrino mass
 in the Gustafsson-No-Rivera model.
 }
 \label{fig:GNR}
\end{minipage}
\end{figure}

 The {\bf Aoki-Kanemura-Seto model}~(AKS model)~\cite{AKS}
is also a 3-loop model of the Majorana neutrino mass
with a dark matter candidate.
 Instead of $s_1^+$ in the KNT model,
$s^0$ with the $Z_2$-odd parity
and the second $\SU(2)_L$-doublet scalar field $\Phi_2$
with the $Z_2$-even parity are utilized
in the 3-loop diagram~(Fig.~\ref{fig:AKS}).
 Since both of two $\SU(2)_L$-doublet scalar fields are
$Z_2$-even ones,
the scalar potential in this model
has the CP-violating phases
which can be utilized for the electro-weak baryogenesis.
 Simplification with $m_e^{} = 0$ is not allowed,
and then the Yukawa interaction of $s_2^-$
is dominated by the one with $e_R^{}$.

 The 3-loop diagram~(Fig.~\ref{fig:GNR})
in the {\bf Gustafsson-No-Rivera model}~(GNR model)~\cite{Gustafsson:2012vj}
involves a dark matter candidate%
~($\text{Re}(\phi_2^0)$ or $\text{Im}(\phi_2^0)$) and the $W$ boson.
 The structure of the neutrino mass matrix
is simply determined by
a unique matrix of new Yukawa coupling constants
and a known diagonal matrix of charged lepton masses.
 Inversely,
the structure of the new Yukawa matrix
is directly constrained by the neutrino oscillation data.
 Notice that
$m_e$ cannot be ignored in the loop diagram in the GNR model
in contrast with the cases in the Zee model and the ZB model.



\begin{figure}[t]
\begin{minipage}{0.45\hsize}
 \vspace*{8mm}
 \includegraphics[scale=0.45]{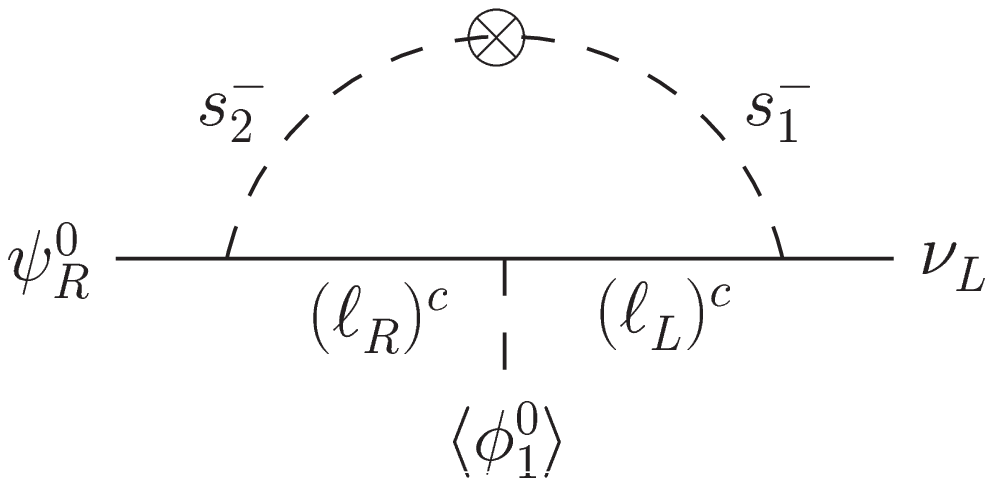}
 \vspace*{-2mm}
 \caption{
 The loop diagram for the Dirac mass of the neutrino
 in the Nasri-Moussa model.
 }
 \label{fig:NM}
\end{minipage}
\qquad
\begin{minipage}{0.45\hsize}
 \includegraphics[scale=0.45]{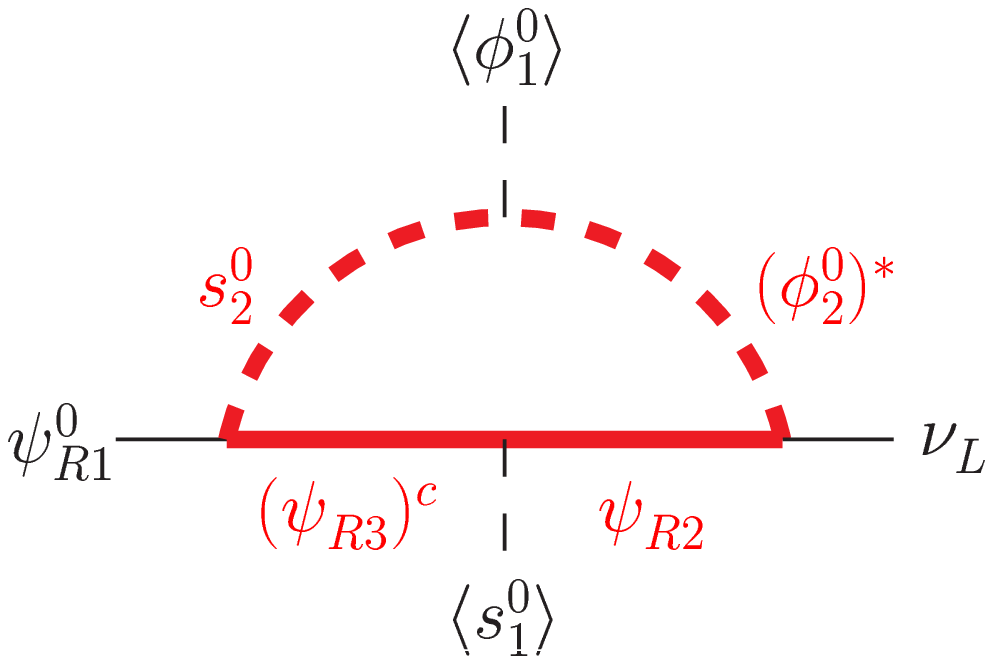}
 \vspace*{-2mm}
 \caption{
 The loop diagram for the Dirac neutrino mass
 in the Gu-Sarkar model.
 }
 \label{fig:GS}
\end{minipage}
\end{figure}

 Dirac masses for neutrinos can be also radiatively generated
by using a softly-broken symmetry~(e.g., $Z_2$, a global $\U(1)$)
which forbids some tree level interactions.
 In the {\bf Nasri-Moussa model}~(NM model)~\cite{Nasri:2001ax}%
~(See also Ref.~\cite{Kanemura:2011jj}),
$\psi_R^0$ and $s_2^+$
are introduced as $Z_2$-odd fields
while another scalar field $s_1^+$
is a $Z_2$-even one.
 An Yukawa interaction $\overline{L} \epsilon \Phi_1^\ast \psi_R^0$
is forbidden by the $Z_2$ symmetry.
 However,
the Yukawa interaction arises at the 1-loop level~(Fig.~\ref{fig:NM})
because the $Z_2$ symmetry is softly-broken
by the $m_{12}^2 s_1^+ s_2^-$ term.
 Then, neutrinos acquire Dirac masses,
and $\psi_R^0$ become the right-handed neutrinos.

In the {\bf Gu-Sarkar model}~(GS model)~\cite{Gu:2007ug},
dark matter candidates are involved
in the 1-loop diagram~(Fig.~\ref{fig:GS})
for Dirac neutrino masses.
 Although fermions $\psi_{R1}^0$ and $(\psi_{R3}^0)^c$
and a complex scalar field $s_1^0$
are singlet under the gauge group of the standard model,
they have a common charge of a $\U(1)^\prime$ gauge symmetry
which forbid Yukawa interactions
$\overline{L} \epsilon \Phi_1^\ast \psi_{R1}^0$ at the tree level.
 The $\U(1)^\prime$ gauge symmetry is spontaneously broken
by a vacuum expectation value~(VEV) of $s_1^0$.
 On the other hand,
an unbroken $Z_2$ symmetry is imposed
such that $\psi_{R2}^0$, $\psi_{R3}^0$,
$s_2^0$, and $\Phi_2$ are odd under the symmetry.
 The conservation of the lepton number
is also imposed such that
$\psi_{R1}^0$, $\psi_{R2}^0$, and $(\psi_{R3}^0)^c$
have a common lepton number $1$
while three new scalar fields have no lepton number;
 the lepton number conservation is necessary to forbids
the diagram in Fig.~\ref{fig:Ma}
with the Majorana mass term of $\psi_{R2}^0$.
 Dirac neutrinos are made from $\nu_L^{}$ and $\psi_{R1}^0$.



\begin{figure}[t]
\begin{minipage}{0.45\hsize}
 \vspace*{1mm}
 \includegraphics[scale=0.45]{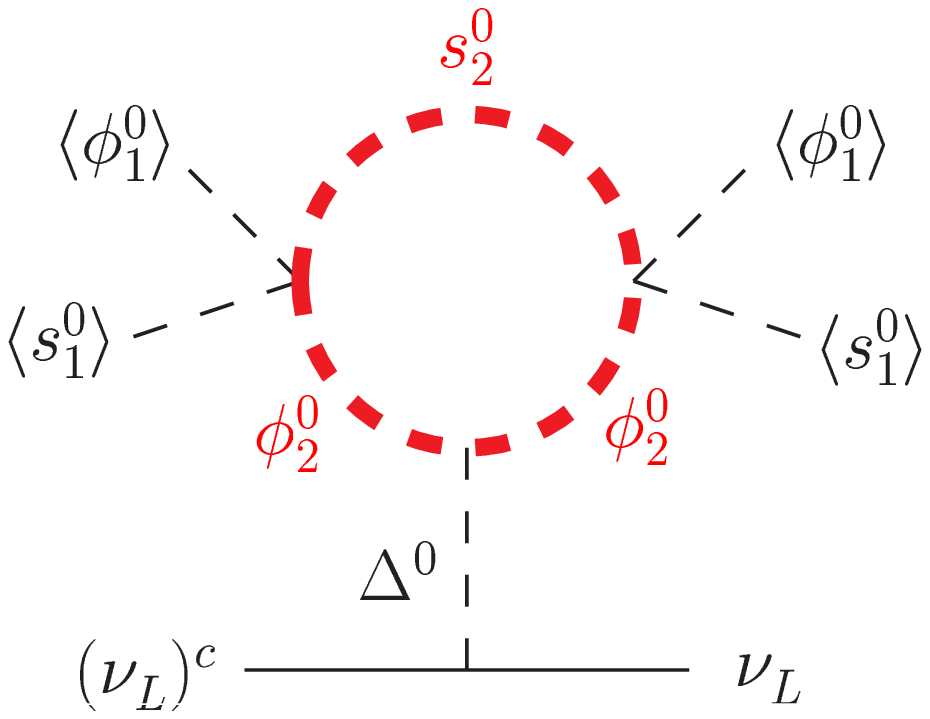}
 \vspace*{-2mm}
 \caption{
 The loop diagram for the VEV of an $\SU(2)_L$-triplet scalar field
 in the Kanemura-Sugiyama model.
 }
 \label{fig:KS}
\end{minipage}
\qquad
\begin{minipage}{0.45\hsize}
\begin{center}
 \includegraphics[scale=0.45]{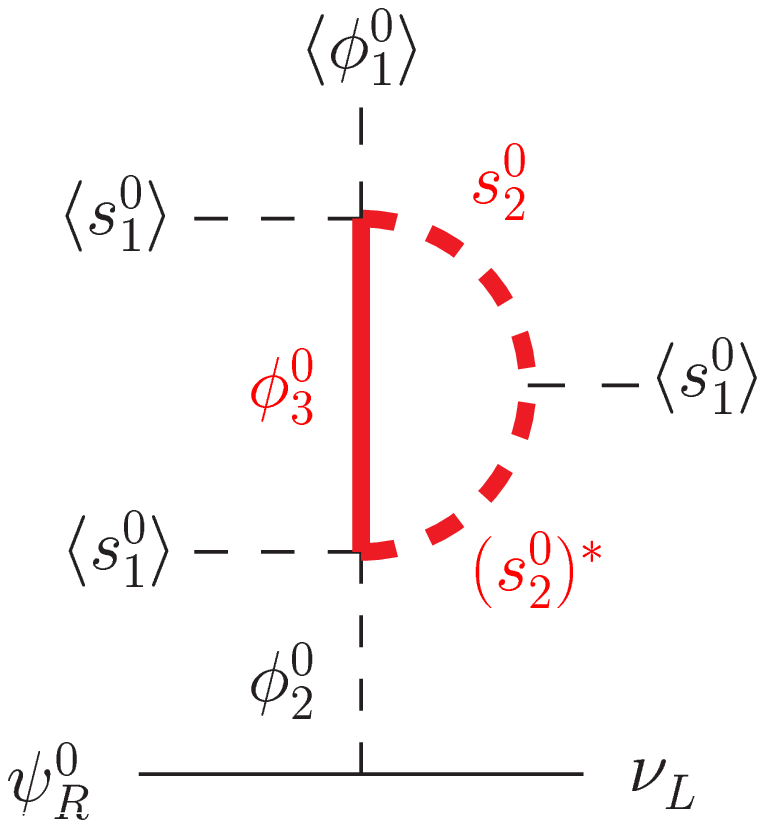}
 \vspace*{-2mm}
 \caption{
 The loop diagram for the VEV of an $\SU(2)_L$-doublet scalar field
 which couples only with the neutrino
 in the Kanemura-Matsui-Sugiyama model.
 }
 \label{fig:KMS}
\end{center}
\end{minipage}
\end{figure}

 In models shown above,
interactions between neutrinos and scalar fields
are induced at the loop level.
 In contrast,
the {\bf Kanemura-Sugiyama model}~(KS model)~\cite{Kanemura:2012rj}
is an extension of the Higgs triplet model~(HTM)~\cite{Ref:HTM}
such that a VEV of an $\SU(2)_L$-triplet scalar field $\Delta$
arises at the 1-loop level
while a Yukawa interaction of Majorana neutrinos
with the triplet scalar field exists at the tree level%
~(Fig.~\ref{fig:KS}).
 Scalar fields $s_2^0$ and $\Phi_2$ are
introduced as $Z_2$-odd ones
while $s_1^0$ is a $Z_2$-even field.
 A lepton number $-1$ is assigned to $\Phi_2$ and $s_1^0$,
and a VEV of $s_1^0$ spontaneously breaks
the lepton number conservation
without breaking the $Z_2$ symmetry
which stabilizes a dark matter candidate.
 The direct relation between the Yukawa matrix with $\Delta$
and the neutrino mass matrix remain the same
as the one in the HTM\@.

 The {\bf Kanemura-Matsui-Sugiyama model}%
~(KMS model)~\cite{Kanemura:2013qva}
is an extension of a version of the two Higgs doublet model
where the second $\SU(2)_L$-doublet scalar field $\Phi_2$
has the Yukawa interaction only with neutrinos~($\nu$THDM)~\cite{Ref:nuTHDM-1}.
 The VEV of $\Phi_2$ in the KMS model
is obtained via a 1-loop diagram in Fig.~\ref{fig:KMS}.
 A global $\U(1)$ symmetry is imposed such that
charges of $\psi_R^0$, $s_1^0$, and $\Phi_2$
are $3$, $1$, and $3$, respectively.
 Fields which exist in the standard model
have no charge for the global $\U(1)$ symmetry,
and Yukawa interaction between $\psi_R^0$ and $\Phi_1$ is forbidden.
 The $\U(1)$ symmetry is spontaneously broken by a VEV of $s_1^0$.
 On the other hand,
$s_2^0$ and $\Phi_3$ have fractional charges $1/2$ and $3/2$,
respectively;
 there appears an accidental unbroken global $\U(1)_D$ symmetry
under which $s_2^0$ and $\Phi_3$ has the same charge,
which stabilizes a dark matter candidate.


\section{Phenomenology}


 For the lepton flavor violation in charged lepton decays,
it would be naively expected
that three-body decays $\ell \to \overline{\ell}_1 \ell_2 \ell_3$
are rarer than two-body decays $\ell \to \ell^\prime \gamma$.
 However,
$\ell \to \overline{\ell}_1 \ell_2 \ell_3$
can be caused at the tree level
while $\ell \to \ell^\prime \gamma$
are always given in the loop level.
 Such tree level $\ell \to \overline{\ell}_1 \ell_2 \ell_3$
are given by the FCNC in the Zee model
or mediated by a doubly-charged scalar particle
which exists in the ZB, the GNR, and the KS models;
$\tau \to \overline{e}\mu\mu$ is dominant
for a benchmark point in the GNR model~\cite{Gustafsson:2012vj},
and the KS model favors $\tau \to \overline{\mu}\mu\mu$,
$\tau \to \overline{\mu}\mu e$, and $\tau \to \overline{\mu}e e$%
~(See e.g., Ref.~\cite{Akeroyd:2009nu} for the Higgs triplet model).
 If some $\ell \to \overline{\ell}_1 \ell_2 \ell_3$ processes are observed
(especially,
in the case without $\ell \to \ell^\prime \gamma$ signal),
these models would be supported.



 A doubly-charged scalar particle
can decay into a pair of same-signed charged leptons,
$H^{--} \to \ell \ell^\prime$.
 Such a particle is involved
in the ZB, the GNR, and the KS models.
 In the ZB model,
it is naively expected that
decay branching ratios
for $s^{--} \to \mu_R^{} \tau_R^{}$ and $\tau_R^{}\tau_R^{}$
are suppressed by $(m_\mu/m_\tau)^2$ and $(m_\mu/m_\tau)^4$,
respectively, in comparison with the ratio for
$s^{--} \to \mu_R^{} \mu_R^{}$%
~(See e.g., Ref,~\cite{AristizabalSierra:2006gb}).
Thus,
$s^{--} \to \tau_R^{} \tau_R^{}$ is not
expected to be observed in the ZB model.
 Similarly to the ZB model,
a matrix $h_{\ell\ell^\prime}^{\text{(GNR)}}$
of Yukawa coupling constants for $s^{--}$ in the GNR model
has a very hierarchical structures
because of the charged lepton masses
in the loop diagram~(Fig.~\ref{fig:GNR}),
$h_{\ell\ell^\prime}^{\text{(GNR)}}
\propto (m_\nu^{})_{\ell\ell^\prime}/( m_\ell^{} m_{\ell^\prime} )$.
 Therefore,
the leptonic decay of $s^{--}$ prefers to involve an electron.
 A decay $s^{--} \to e_R^{}\tau_R^{}$ is dominant
for a scenario where
both of $|(m_\nu^{})_{ee}|$ and $|(m_\nu^{})_{e\mu}|$
are assumed to be negligible,
and a benchmark values of parameters for the scenario
is shown in Ref.~\cite{Gustafsson:2012vj}.
 In the KS model,
predictions for $\Delta^{--} \to \ell_L^{} \ell_L^\prime$
are the same as those in the Higgs triplet model%
~(See e.g., Ref.~\cite{Akeroyd:2007zv}).



 Singly-charged scalar particles are
involved in all models in Table~\ref{tab:models}.
 Mixings between scalar particles are ignored
in most of discussion below for simplicity.
 Since we do not observe flavors of neutrinos in $H^-$ decays,
let us define branching ratios
$\BR(H^- \to \ell \nu)
\equiv \sum_{\ell^\prime} \BR(H^- \to \ell \nu_{\ell^\prime}^{})$.
%
%
 Flavor structures of
$\BR(\phi_2^- \to \ell_R^{} \overline{\nu_L^{}})$
in the Zee model
and $\BR(s_2^- \to \ell_R^{} \psi_R^0)$
in the NM model are arbitrary.
%
%
 For an antisymmetric matrix $f_{\ell\ell^\prime}^{\text{(ZM)}}$
of Yukawa coupling constants for $s^-$ in the Zee model,
a simplification with $m_e^{} = m_\mu^{} = 0$
results in $|f_{\mu\tau}^{\text{(ZM)}}|^2 \ll |f_{e\tau}^{\text{(ZM)}}|^2$.
 Then,
the Zee model predicts
$\BR(s^- \to e_L^{} \nu_L^{}) :
(\BR(s^- \to \mu_L^{} \nu_L^{}) + \BR(s^- \to \tau_L^{} \nu_L^{}))
\simeq 1 : 1$.
%
%
 Decay branching ratios
$\BR(s^- \to \ell_L^{} \nu_L^{})$
in the ZB model for Majorana mass terms
and $\BR(s_1^- \to \ell_L^{} \nu_L^{})$
in the NM model for Dirac mass terms
have a common flavor structure.
 These models predict
$\BR(H^- \to e_L^{} \nu_L^{}) : \BR(H^- \to \mu_L^{} \nu_L^{})
: \BR(H^- \to \tau_L^{} \nu_L^{}) \simeq 2 : 5 : 5$
for the so-called normal mass ordering%
~($m_1 < m_3$ where $m_i^{}$ denote neutrino masses)
and $2 : 1 : 1$ for the so-called inverted mass ordering~($m_3 < m_1)$.
%
%
 In the AKS model,
the second $\SU(2)_L$-doublet field
couples only with leptons,
and its VEV gives charged lepton masses.
 Therefore,
the leptonic decay of $\phi_2^-$ is
dominated by the decay into $\tau_R^{}$.

 There exists a singly-charged scalar particle
from an $\SU(2)_L$-doublet field
which is $Z_2$-odd~(or charged under a global $\U(1)$)
in the Ma, the GNR, the KS, the GS, and the KMS models.
%
%
 In the GNR, the KS, and the KMS models,
$\phi_2^-$ dominantly decays into $W^-$.
 Then,
the decay of the $\phi_2^-$
gives charged leptons
with the equal ratio $e : \mu : \tau = 1 : 1 : 1$.
%
%
 This is also the case in the Ma and the GS models
if the weak decay of $\phi_2^-$ is dominant.
 If $\phi_2^- \to \ell_L^{} \overline{\psi_R^0}$
is dominant in these two models,
we do not have clear predictions
on the flavor structure.

\begin{figure}[t]
 \includegraphics[scale=0.3]{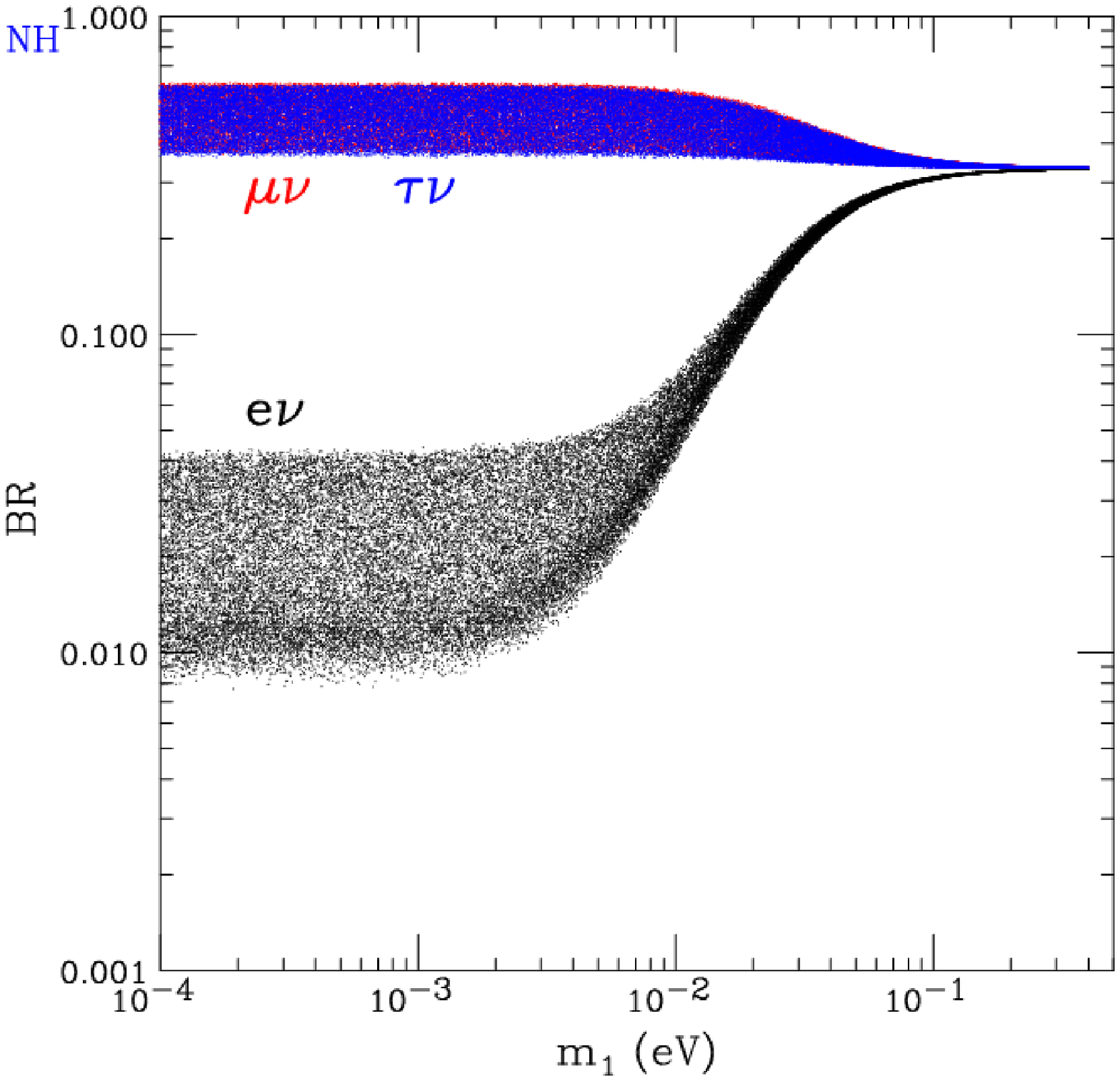} \quad
 \includegraphics[scale=0.3]{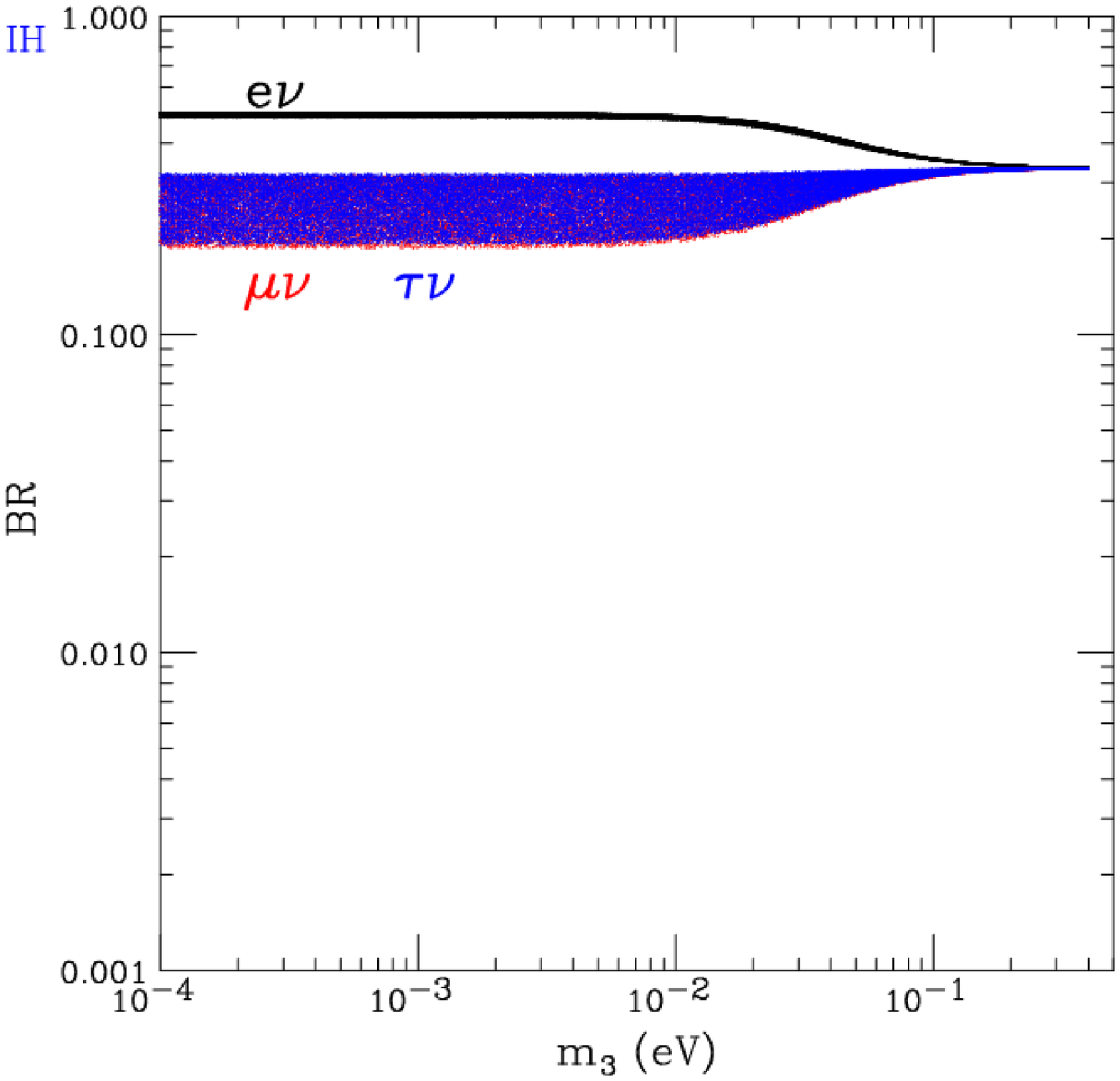}
 \vspace*{-2mm}
 \caption{
 Leptonic decays of a charged Higgs boson in the HTM
 for $m_1 < m_3$~(left) and $m_3 < m_1$~(right),
 where $m_i$ are neutrino mass eigenstates.
 These plots are taken from Ref.~\cite{Perez:2008ha},
which are the same in the KS and the KMS models.
 }
 \label{fig:decay}
\end{figure}

 The KNT, the AKS, and the GNR models
contain a singly-charged $\SU(2)_L$-singlet scalar particle
with the $Z_2$-odd parity.
%
%
 In the KNT model,
the Yukawa interaction of $s_2^-$ with $\tau_R^{}$
would be suppressed by $m_\mu^{}/m_\tau^{}$
in comparison with that with $\mu_R^{}$
(See e.g., Ref.~\cite{Cheung:2004xm}
for benchmark values of parameters)
similarly to the expected hierarchy
in Yukawa coupling constants for $s^{--}$ in the ZB model.
 Therefore,
branching ratio for $s_2^- \to \tau_R^{} \psi_R^0$
becomes too tiny to be measured.
%
%
 In the AKS model,
$s^-$ can decay as $s^- \to \phi_2^- s^0$
followed by the decay of $\phi_2^-$ into $\tau_R^{}$
for parameter sets in Ref.~\cite{AKS}.
 If $s^- \to \ell_R^{} \psi_R^0$ is kinematically possible,
$s^-$ dominantly decays into an electron.
%
%
 The $s^-$ in the GNR model
decays into $W$ through
the mixing between $s^-$ and $\phi_2^-$,
and the ratio of produced charged leptons
is the same as that for the $\phi_2^-$ decay,
$e : \mu : \tau = 1 : 1 : 1$.
%
%
 Prediction about leptonic decays of singly-charged Higgs bosons
in the KS and KMS models
are the same as those in the HTM model
and the $\nu$THDM model, respectively,
because there is no extension for Yukawa interactions.
 Figure~\ref{fig:decay}~(taken from Ref.~\cite{Perez:2008ha})
shows the prediction in the HTM,
which is also the one in the $\nu$THDM~\cite{Ref:nuTHDM-2};
 the Fig.~\ref{fig:decay}
can be used for both of the KS and KMS models.

\subsection{Summary}

 New Higgs bosons are introduced in radiative neutrino mass models
where neutrino masses are generated at the loop level,
and it is not necessary for these bosons to be very heavy.
 Since their Yukawa interactions relate to
the structure of neutrino mass matrix
which is constrained by the neutirno oscillation data,
we have predictions about
$\ell \to \overline{\ell}_1\ell_2\ell_3$
and leptonic decays of singly and doubly-charged Higgs bosons.
 Such predictions can be used to test these models.
 We hope that some signal of such processes are observed in near future,
which would drive us to meet again in Toyama.

%

\begin{acknowledgments}
 I would like to thank
all participants in this workshop for coming
in spite of poor weather with heavy snow.
 I enjoyed all presentations and fruitful discussion.
\end{acknowledgments}

\bigskip 

\end{document}